\documentclass[a4paper,11pt]{article}
\usepackage{amsfonts}
\usepackage{amssymb}
\usepackage{graphicx}
\usepackage{amsmath}

\DeclareFontFamily{U}{rsf}{}
\DeclareFontShape{U}{rsf}{m}{n}{
  <5> <6> rsfs5 <7> <8> <9> rsfs7 <10-> rsfs10}{}
\DeclareMathAlphabet\Scr{U}{rsf}{m}{n} \makeatletter
\@addtoreset{equation}{section} \makeatother

\begin{document}

\begin{titlepage}

\begin{flushright}
Imperial/TP/2016/mjd/3 \\
CERN-TH-2016-225 \\
DFPD/2016/TH/18
\end{flushright}

\vskip 1.0 cm
\begin{center}  {\huge{\bf $D=3$ Unification \\\vskip 0.4 cm of Curious Supergravities}}

\vskip 1.0 cm

{\Large {\bf M. J. Duff$^{1,2}$}, {\bf S. Ferrara$^{3,4,5}$} and {\bf A. Marrani$^{6,7,3}$}}

\vskip 1.0 cm

$^1${\sl Theoretical Physics, Blackett Laboratory, Imperial College London,\\ London SW7 2AZ, United Kingdom}\\
\texttt{m.duff@imperial.ac.uk}

\vskip 0.2 cm

$^2${\sl Mathematical Institute University of Oxford
Andrew Wiles Building\\
Woodstock Road, Radcliffe Observatory Quarter,\\ Oxford, OX2 6GG, United Kingdom}

\vskip 0.2 cm

$^3${\sl Theoretical Physics Department, CERN, \\ CH-1211 Geneva, Switzerland}\\
\texttt{sergio.ferrara@cern.ch}

\vskip 0.2 cm

$^4${\sl INFN - Laboratori Nazionali di Frascati, \\ Via Enrico Fermi 40, I-00044 Frascati, Italy}\\

\vskip 0.2 cm

$^5${\sl Department of Physics and Astronomy,\\and Mani L. Bhaumik Institute for Theoretical Physics,\\UCLA, Los Angeles CA 90095-1547, USA}\\

\vskip 0.2 cm

$^6${\sl Museo Storico della Fisica e Centro Studi e
Ricerche ``Enrico Fermi" \\ Via Panisperna 89A, I-00184, Roma, Italy}\\

\vskip 0.2 cm

$^7${\sl Dipartimento di Fisica e Astronomia ``Galileo
Galilei'', Universit\`a di Padova
\\ and INFN, Sez. di Padova \\
Via Marzolo 8, I-35131 Padova, Italy}
\\ \texttt{Alessio.Marrani@pd.infn.it}\\

\vskip 0.2 cm

 \end{center}

 \vskip 1.0 cm

\begin{abstract}
We consider the dimensional reduction to $D=3$ of four maximal-rank
supergravities which preserve minimal supersymmetry in $D=11$, $7$, $5$ and $%
4$. Such \textit{\textquotedblleft curious"} theories were investigated some
time ago, and the four-dimensional one corresponds to an $\mathcal{N}=1$
supergravity with $7$ chiral multiplets spanning the seven-disk manifold.
Recently, this latter theory provided cosmological models
for $\alpha $-attractors, which are based on the disk geometry with possible
restrictions on the parameter $\alpha $. A unified picture emerges in $D=3$, where the Ehlers group of General
Relativity merges with the $S$-, $T$- and $U$- dualities of the $D=4$ parent
theories.
\end{abstract}
\vspace{24pt} \end{titlepage}




\newpage

\section{\label{Intro}Introduction}

Among compactifications of $D=11$ supergravity on a $7$-manifold to $D=4$,
an interesting $\mathcal{N}=1$ theory emerges, whose spectrum consists of
seven chiral (Wess-Zumino) multiplets living in the seven-disk manifold%
\begin{equation}
\left[ \frac{SL(2,\mathbb{R})}{U(1)}\right] ^{\otimes 7}.  \label{0}
\end{equation}%
This theory, proposed in \cite{Duff-Ferrara-curious} has some peculiar
properties. It is the smallest member of a family of four \textit{%
\textquotedblleft curious"} supergravities, defined in $D=(11,7,5,4)$
dimensions, having a scalar manifold of (maximal) rank $(0,4,6,7)$,
respectively, and endowed with a minimal number $\nu $ of supersymmetries in
the corresponding dimensions, $\nu =(32,16,8,4)$, respectively. Such
theories couple naturally to supermembranes and admit these membranes as
solutions. In \cite{Ferrara-Kallosh} the seven-disk manifold (\ref{0}) was
considered as providing possible restrictions on the parameter $\alpha $ of
the cosmological $\alpha $-attractors models for inflation, depending on the
embeddings of the single one-disk into (\ref{0}).

When compactified on a $7$-manifold $X^{7}$ with Betti numbers $\left(
b_{0},b_{1},b_{2},b_{3}\right) =\left( b_{7},b_{6},b_{5},b_{4}\right) $, the
number of fields of spin $s=(2,3/2,1,1/2,0)$ in the resulting $D=4$
supergravity is given by $%
n_{s}=(b_{0},b_{0}+b_{1},b_{1}+b_{2},b_{2}+b_{3},2b_{3})$, and we may
loosely associate Betti numbers with any supergravity with $n_{s}$ fields of
spin $s$, whether or not manifolds with these Betti numbers actually exist.
We may then define a generalized mirror transformation \cite%
{Duff-Ferrara-curious}
\begin{equation}
(b_{0},b_{1},b_{2},b_{3})\rightarrow (b_{0},b_{1},b_{2}-\rho /2,b_{3}+\rho
/2),
\end{equation}%
%
%
%
%
%
%
%
under which
\begin{equation}
\rho \left( X^{7}\right) :=\sum_{k=0}^{7}\left( -1\right) ^{k+1}\left(
k+1\right) b_{k}=7b_{0}-5b_{1}+3b_{2}-b_{3},
\end{equation}%
changes sign:
\begin{equation}
\rho \rightarrow -\rho
\end{equation}%
(In the special case $b_{1}=0$, $\rho $ reversal reduces to the reflection
symmetry of $G_{2}$ manifolds defined by Joyce \cite{Joyce-16,Joyce-17}).
Generalised self-mirror theories are here defined to be those for which $%
\rho $ vanishes. Under further toroidal compactification to $D=4$, the four
curious supergravities have $\mathcal{N}=8,4,2,1$ supersymmetries and Betti
numbers $(b_{0},b_{1},b_{2},b_{3})=(1,\mathcal{N}-1,n,3n-5\mathcal{N}+12)$
and thus are all self-mirror. (The $\mathcal{N}=2$ theory is just the
self-mirror $stu$ model \cite{STU}.)

Similarly, we may define a generalized mirror transformation for $6$%
-manifolds $X^{6}$ \cite{Duff-Ferrara-curious} with Betti numbers $\left(
c_{0},c_{1},c_{2},c_{3}\right) =\left( c_{6},c_{5},c_{4},c_{3}\right) $ :
\begin{equation}
(c_{0},c_{1},c_{2},c_{3})\rightarrow (c_{0},c_{1},c_{2}-\chi /2,c_{3}+\chi )
\end{equation}%
under which
\begin{equation}
c\left( X^{6}\right) :=\sum_{k=0}^{6}\left( -1\right)
^{k}c_{k}=2c_{0}-2c_{1}+2c_{2}-c_{3}
\end{equation}%
changes sign:
\begin{equation}
\chi \rightarrow -\chi
\end{equation}%
(In the special case $c_{1}=0$, $\chi $ reversal reduces to ordinary mirror
symmetry of Calabi-Yau \cite{CY}). Generalised self-mirror theories are here
defined to be those for which $\chi $ vanishes. In the special case $%
X^{7}=X^{6}\times S^{1}$, $\rho =\chi $ and the two symmetries
coincide.\medskip

Given the unusual properties and possible cosmological applications of these
curious supergravities, in the present note we give a $D=3$ three-way
unified picture in terms of

\textbf{1)} compactifications of $M$-theory in terms of toroidal moduli;

\textbf{2)} dimensional reduction of the four curious supergravities $%
D=(11,7,5, 4)$ to $D=3$;

\textbf{3)} dimensional reduction of 4 curious supergravities in $D=4$ to $%
D=3$. In particular, the resulting $\mathcal{N}=2$, $D=3$ supergravity has
the scalar manifold given by the eight-disk manifold%
\begin{equation}
\left[ \frac{SL(2,\mathbb{R})}{U(1)}\right] ^{\otimes 8},  \label{1}
\end{equation}%
which can be regarded as the unification of $S$-, $T$- and $U$- dualities of
the $\mathcal{N}=1$, $D=4$ corresponding theory mentioned above, augmented
by the disk manifold $\frac{SL(2,\mathbb{R})_{Ehlers}}{U(1)}$ pertaining to
the $D=4$ Ehlers group $SL(2,\mathbb{R})_{Ehlers}$.\medskip

The paper is organized as follows.

In Sec. \ref{Eight-Disc} we recall the embedding of $\left[ SL(2,\mathbb{R})%
\right] ^{\otimes 8}$ into $E_{8(8)}$. In Sec. \ref{M-Theory} we give an
interpretation of the four curious supergravities in terms of sequential
reductions of $M$-theory on an eight-manifold with only toroidal moduli of $%
T^{8}$, $T^{4}\times T^{4}$, and $T^{2}\times T^{2}\times T^{2}\times T^{2}$
(\textit{\textquotedblleft }$M$\textit{-theoretical path"}). Then, in Sec. %
\ref{Ehlers} we consider the so-called \textit{\textquotedblleft Ehlers path"%
}, by compactifying these theories from $D=4$ to $D=3$. Finally, Sec. \ref%
{Conclusion} contains some concluding remarks.

\section{\label{Eight-Disc}$E_{8(8)}$ and the Eight-Disk Manifold}

Almost all exceptional Lie algebras $\mathcal{E}$ enjoy a rank-preserving
(generally non--maximal nor symmetric) embedding of the type%
\begin{equation}
\mathcal{E}\supset \left[ \mathfrak{sl}(2)\right] ^{\oplus r},~r:=\text{rank}%
(\mathcal{E}).
\end{equation}%
This holds for $\mathcal{E}=\mathfrak{e}_{8},\mathfrak{e}_{7},\mathfrak{f}%
_{4},\mathfrak{g}_{2}$, with $r=8,7,4,2$, respectively. The unique exception%
\footnote{%
It should be here pointed that $\mathfrak{e}_{6}$ stands on its own among
exceptional Lie algebras for \textit{at least} another reason : it is the
unique exceptional Lie algebra which does not embed maximally its principal
(Kostant's) $\mathfrak{sl}(2)_{P}$ \cite{Kostant} algebra. Indeed, while all
Lie algebras maximally embed $\mathfrak{sl}(2)_{P}$ ($\mathfrak{e}_{8}$ and $%
\mathfrak{e}_{7}$ actually maximally embed three and two $\mathfrak{sl}(2)$%
's , respectively), $\mathfrak{e}_{6}$ embeds its $\mathfrak{sl}(2)_{P}$
through the chain of maximal embeddings $\mathfrak{e}_{6}\supset \mathfrak{f}%
_{4}\supset \mathfrak{sl}(2)_{P}$ (in other words, $\mathfrak{e}_{6}$
"inherits" the $\mathfrak{sl}(2)_{P}$ of $\mathfrak{f}_{4}$).} is provided
by the rank-$6$ exceptional algebra $\mathfrak{e}_{6}$, which embeds only $%
\left[ \mathfrak{sl}(2)\right] ^{\oplus 4}$, and not $\left[ \mathfrak{sl}(2)%
\right] ^{\oplus 6}$.

In the following treatment, we will focus on the maximally non-compact (%
\textit{i.e.}, split) real form $\mathfrak{e}_{8(8)}$ of $\mathfrak{e}_{8}$,
considering it at the Lie group level ($E_{8(8)}\supset \left[ SL(2,\mathbb{R%
})\right] ^{\otimes 8}$), in the context of $D=3$ supergravity theories.

More specifically, starting from\footnote{$E_{8(8)}$ belongs to the
so-called \textit{exceptional }$E_{n(n)}$\textit{-sequence} \cite{CJ-1, CJ-2}
of symmetries of maximal supergravities in $11-n$ dimensions.} $E_{8(8)}$ we
will analyze two paths yielding the same $\mathcal{N}=2$, $D=3$ supergravity
theory\footnote{%
For a thorough analysis of the geometric structure of scalar manifolds of $%
D=3$ supergravity theories, see \cite{Tollsten}.}, coupled to $8$ matter
multiplets, whose scalars coordinatize the completely factorized rank%
\footnote{%
The \textit{rank} of a manifold is defined as the maximal dimension (in $%
\mathbb{R}$) of a flat (\textit{i.e.}, with vanishing Riemann tensor),
totally geodesic submanifold (see e.g. $\S $6, page 209 of \cite{ref-4}).}-$%
8 $ Hodge-K\"{a}hler symmetric, \textit{eight-disk manifold} (\ref{1}).

\section{\label{M-Theory}The $M$-Theory Path}

The first path starts from $M$-theory (or, more appropriately, $\mathcal{N}%
=1 $, $D=11$ supergravity), and performs iterated compactifications on tori $%
T^{8}$, $T^{4}\times T^{4}$, and on $T^{2}\times T^{2}\times T^{2}\times
T^{2}$; this corresponds to the following chain of maximal and symmetric
embeddings:%
\begin{eqnarray}
E_{8(8)} &\supset &SO(8,8)  \label{2} \\
&\supset &SO(4,4)\times SO(4,4)  \label{2-bis} \\
&\supset &\left[ SO(2,2)\right] ^{\otimes 4}\cong \left[ SL(2,\mathbb{R})%
\right] ^{\otimes 8}.  \label{2-tris}
\end{eqnarray}

Each step of this chain has an interpretation in terms of truncations of the
massless spectrum of $M$-theory dimensionally reduced to $D=3$, such as to
preserve $\mathcal{N}=16,8,4,2$ local supersymmetries. As we discuss below,
the last three are obtained keeping only the geometric moduli of the tori $%
T^{8}$, $T^{4}\times T^{4}$ and $T^{2}\times T^{2}\times T^{2}\times T^{2}$,
respectively. It is worth here recalling that the classical moduli space of
a d-dimensional torus is ($I,J=1,...,d$)%
\begin{equation}
M_{d}:=\mathbb{R}^{+}\times \frac{SL(d,\mathbb{R})}{SO(d)},~\text{spanned~by~%
}g_{IJ}=g_{(IJ)}\text{,}
\end{equation}%
whereas the quantum one (in a stringy sense) reads%
\begin{equation}
\mathcal{M}_{d}:=\frac{SO(d,d)}{SO(d)\times SO(d)},~\text{spanned~by~}%
g_{IJ}=g_{(IJ)}\text{~and~}B_{IJ}=B_{[IJ]}.
\end{equation}

The first, starting step of the $M$\textit{-theoretical path} (\ref{2})-(\ref%
{2-tris}) corresponds to\footnote{%
"$B$" and "$F$" denote the number of bosonic and fermionic massless degrees
of freedom throughout.} :%
\begin{equation}
M\text{-theory}\overset{T^{8}~\text{(geom}+\text{non-geom)}}{\longrightarrow
}\underset{\left( B,F\right) =\left( 128,128\right) }{\mathcal{N}=16,D=3}:%
\frac{E_{8(8)}}{SO(16)},  \label{128}
\end{equation}%
namely a compactification retaining \textit{both} geometric ($g_{IJ}$, $%
A_{\mu IJ}$; ) and non-geometric ($g_{\mu I}$, $A_{IJK}$) moduli of $T^{8}$,
down to maximal supergravity in $D=3$ \cite{Marcus} ($I,J,K=1,...,8$, and $%
\mu =0,1,2$); note that the $128$ bosonic massless degrees of freedom can be
organized in $SO(8)$ irreprs. as follows :%
\begin{equation}
\underset{\mathbf{35}+\mathbf{1}}{g_{IJ}},~\underset{\mathbf{28}}{A_{\mu IJ}}%
,~\underset{\mathbf{8}}{g_{\mu I}},~\underset{\mathbf{56}}{A_{IJK}},
\end{equation}%
where the $1$-form $A_{\mu IJ}=A_{\mu \lbrack IJ]}$ (playing the role of the
\textquotedblleft $M$-theoretical $B$-field") gets then dualized to scalar
fields $A_{IJ}$ in $D=3$.

The next step corresponds to the first, maximal and symmetric embedding (\ref%
{2}), which amounts to retaining only the geometric moduli of $T^{8}$ (%
\textit{i.e.}, to setting $g_{\mu I}=0=A_{IJK}$ in the bosonic sector), thus
giving rise upon compactification to half-maximal supergravity coupled to $%
n=8$ matter multiplets in $D=3$ :%
\begin{equation}
M\text{-theory}\overset{T^{8}~\text{(geom)}}{\longrightarrow }\underset{%
\left( B,F\right) =\left( 64,64\right) }{\mathcal{N}=8,D=3,n=8}~:\frac{%
SO(8,8)}{SO(8)\times SO(8)}.  \label{64}
\end{equation}

The subsequent maximal and symmetric embedding (\ref{2-bis}) corresponds to
a compactification on $T^{4}\times T^{4}$ retaining only the corresponding
geometric moduli ($i,j=1,...,4$, and $i^{\prime },j^{\prime }=5,...,8$):%
\begin{equation}
g_{ij},~A_{\mu ij},~g_{i^{\prime }j^{\prime }},~A_{\mu i^{\prime }j^{\prime
}},
\end{equation}%
thus giving rise to the following $\mathcal{N}=4$, $D=3$ supergravity model :%
\begin{equation}
M\text{-theory}\overset{T^{4}\times T^{4}~\text{(geom)}}{\longrightarrow }%
\underset{\left( B,F\right) =\left( 32,32\right) }{\mathcal{N}=4,D=3,n=8}:%
\frac{SO(4,4)}{SO(4)\times SO(4)}\times \frac{SO(4,4)}{SO(4)\times SO(4)}.
\label{32}
\end{equation}

The last step is given by the maximal and symmetric embedding (\ref{2-tris}%
), corresponding to a compactification on $T^{2}\times T^{2}\times
T^{2}\times T^{2}$ retaining only the related geometric moduli%
\begin{equation}
g_{11},g_{12},g_{22},A_{\mu 12},~~g_{33},g_{34},g_{44},A_{\mu
34},~~g_{55},g_{56},g_{66},A_{\mu 56},~~g_{77},g_{78},g_{88},A_{\mu 78},
\end{equation}%
thus giving rise to the $\mathcal{N}=2$, $D=3$ supergravity model whose
scalar manifold is given by the eight-disk manifold (\ref{1}):%
\begin{equation}
M\text{-theory}\overset{T^{2}\times T^{2}\times T^{2}\times T^{2}~\text{%
(geom)}}{\longrightarrow }\underset{\left( B,F\right) =\left( 16,16\right) }{%
\mathcal{N}=2,D=3}:\left[ \frac{SL(2,\mathbb{R})}{U(1)}\right] ^{\otimes 8}.
\label{16}
\end{equation}

Some comments are in order.

\begin{enumerate}
\item All symmetric scalar manifolds in (\ref{128}), (\ref{64}), (\ref{32})
and (\ref{16}) have rank $8$, as a consequence of the fact that all
embeddings of the chain (\ref{2})-(\ref{2-tris}) are rank-preserving.

\item The theories (\ref{128}), (\ref{64}), (\ref{32}) and (\ref{16}) are
nothing but the $D=3$ reduction of the four curious supergravities, studied
in \cite{Duff-Ferrara-curious} and mentioned in Sec. \ref{Intro}. These
latter are defined in $D=q+3=11,7,5,4$ Lorentzian space-time dimensions
(with $q:=\dim _{\mathbb{R}}\mathbb{A}=8,4,2,1$, where $\mathbb{A}=\mathbb{O}
$ (octonions), $\mathbb{H}$ (quaternions), $\mathbb{C}$ (complex numbers), $%
\mathbb{R}$ (reals) denote the four Hurwitz division algebras),with scalar
manifolds of rank $0,4,6,7$ respectively. As observed in \cite%
{Duff-Ferrara-curious}, such $\mathcal{N}=8,4,2,1$, $D=4$ curious
supergravities respectively correspond to $\mathcal{N}-1=7,3,1,0$ lines of
the Fano plane, and hence they admit a division algebraic interpretation
consistent with the so-called \textit{"black-hole/qubit" correspondence} (%
\textit{cfr. e.g.} \cite{BH-Qubit-rev-1} for an introduction and a list of
Refs.). By further compactifying them respectively on $T^{8}$, $T^{4}$, $%
T^{2}$, $T^{1}=S^{1}$ down to $D=3$, the rank of the corresponding scalar
manifold (after dualization) increase by $8,4,2,1$, so that all the
resulting $D=3$ theories have rank-$8$ scalar manifolds, as given by (\ref%
{128}), (\ref{64}), (\ref{32}) and (\ref{16}). They have $\mathcal{N}%
=2^{4},2^{3},2^{2},2$ local supersymmetry in $D=3$, with $2^{8}$, $2^{7}$, $%
2^{6}$ and $2^{5}$ total number of massless states, respectively. In this
perspective, the dimensional reduction to $D=3$ provides a \textit{unified
view of the curious supergravities}.
\end{enumerate}

\section{\label{Ehlers}The Ehlers Path}

The second path yielding the $\mathcal{N}=2$, $D=3$ supergravity theory with
scalar manifold (\ref{1}) starts with the so-called \textit{Ehlers embedding}
(\textit{cfr. e.g.} \cite{super-Ehlers}, and Refs. therein) for maximal
supergravity in $D=4\rightarrow D=3$, and then proceeds with a chain of
maximal, symmetric and rank-preserving embeddings which has already been
considered in \cite{Duff-Ferrara-E7,ICL-Rev,Ferrara-Kallosh} :%
\begin{eqnarray}
E_{8(8)} &\supset &E_{7(7)}\times SL(2,\mathbb{R})_{Ehlers}
\label{jazz-pre-2} \\
&\supset &SO(6,6)\times SL(2,\mathbb{R})_{Ehlers}\times SL(2,\mathbb{R})
\label{jazz-2} \\
&\supset &SO(4,4)\times \left[ SL(2,\mathbb{R})\right] ^{\otimes 2}\times
SL(2,\mathbb{R})_{Ehlers}\times SL(2,\mathbb{R})  \label{jazz-3} \\
&\supset &\left[ SL(2,\mathbb{R})\right] ^{\otimes 8}  \label{jazz-4}
\end{eqnarray}

Since this path, which we name \textit{Ehlers path}, starts with a $%
D=4\rightarrow D=3$ dimensional reduction, it is immediate to realize that
the $D=3$ scalar manifolds given in (\ref{128}), (\ref{64}), (\ref{32}) and (%
\ref{16}) are nothing but the dimensional reduction of the $D=4$ cosets of $%
\mathcal{N}=8,4,2,1$ curious supergravities with rank-$7$ scalar manifolds
(after dualization; \textit{cfr.} Table XVIII of \cite{Duff-Ferrara-curious}%
).

While for $\mathcal{N}=8,4,2$ the dimensional reduction $D=4\rightarrow D=3$
is well-known from the study of Maxwell-Einstein systems coupled to
non-linear sigma models (\cite{BGM}, thereby including the $c$-map \cite%
{CFG, Sabha} relating projective special K\"{a}hler manifolds to
quaternionic manifolds), for $\mathcal{N}=1$ the dimensional reduction reads
\begin{equation}
\text{{\small $\left( B,F\right) $\textsc{=}$\left( 16,16\right) $}}:%
\underset{\mathcal{N}=1,D=4,n_{c}=7,n_{v}=0}{\left[ \frac{SL(2,\mathbb{R})}{%
U(1)}\right] ^{\otimes 7}}\longrightarrow \underset{\mathcal{N}=2,D=3,n=8}{%
\left[ \frac{SL(2,\mathbb{R})}{U(1)}\right] ^{\otimes 8}},  \label{4}
\end{equation}%
and it stands on a different footing. Indeed, the $\mathcal{N}=1$, $D=4$
supergravity theory is coupled only to $7$ chiral multiplets, \textit{with
no vectors at all}. Therefore, under (spacelike) dimensional reduction $%
D=4\rightarrow D=3$, the chiral multiplets' scalar manifold (\ref{0}) gets
enlarged only by a further factor manifold $\frac{SL(2,\mathbb{R})_{Ehlers}}{%
U(1)}$, spanned by the axio-dilaton given by the $S^{1}$-radius of
compactification and by the dualization of the corresponding Kaluza-Klein
vector. In other words, the added $\frac{SL(2,\mathbb{R})_{Ehlers}}{U(1)}$
manifold pertains to the two degrees of freedom of the $D=4$ massless
graviton (since in $D=3$ the graviton does not propagate any degree of
freedom) : as mentioned in Sec. \ref{Intro}, the \textit{seven-disk manifold}
(\ref{0}) \cite{Duff-Ferrara-curious, Ferrara-Kallosh} gets enlarged to the
\textit{eight-disk manifold} (\ref{1}) by including the $D=4$ Ehlers group $%
SL(2,\mathbb{R})_{Ehlers}$.

\newpage

Some observations are :

\begin{enumerate}
\item All symmetric scalar manifolds in (\ref{blues-1}), (\ref{blues-2}) and
(\ref{blues-3}) have rank $7$, as a consequence of the fact that all
embeddings of the chain (\ref{jazz-pre-2})-(\ref{jazz-4}) are
rank-preserving.

\item The chain of embeddings (\ref{jazz-pre-2})-(\ref{jazz-4}) has been
used in \cite{Duff-Ferrara-E7} (also \textit{cfr.} \cite{ICL-Rev}) to study
the tripartite entanglement of seven qubits inside $E_{7}$. Moreover, it was
recently exploited in \cite{Ferrara-Kallosh} in order to obtain the $%
\mathcal{N}=1$, $D=4$ theory with $7$ WZ multiplet given in the fourth line
of (\ref{4}).

\item The maximal and symmetric embedding (\ref{jazz-2}) corresponds to the
truncation of maximal $D=4$ supergravity to half-maximal supergravity
coupled to $6$ matter (vector) multiplets :%
\begin{equation}
\underset{\mathcal{N}=8,D=4,~(B,F)=(128,128)}{\frac{E_{7(7)}}{SU(8)}}%
~\longrightarrow ~\underset{\mathcal{N}=4,D=4,n=6,~(B,F)=(64,64)}{\frac{SL(2,%
\mathbb{R})}{U(1)}\times \frac{SO(6,6)}{SO(6)\times SO(6)}}.  \label{blues-1}
\end{equation}

\item The subsequent step (\ref{jazz-3}) corresponds to the truncation of
half-maximal $D=4$ supergravity coupled to $6$ vector multiplets to the $%
\mathcal{N}=2,D=4$ $stu$ model coupled to $4$ hypermultiplets, whose
quaternionic scalars coordinatize the symmetric scalar manifold $\frac{%
SO(4,4)}{SO(4)\times SO(4)}$; since this latter is the $c$-map \cite{CFG} of
the corresponding vector-multiplets' projective special K\"{a}hler manifold $%
\left[ \frac{SL(2,R)}{U(1)}\right] ^{\otimes 3}$, this model is \textit{%
self-mirror} (also \textit{cfr. e.g.} \cite{FM-1}) :%
\begin{equation}
\underset{\mathcal{N}=4,D=4,n=6,~(B,F)=(64,64)}{\frac{SL(2,\mathbb{R})}{U(1)}%
\times \frac{SO(6,6)}{SO(6)\times SO(6)}}~\longrightarrow ~\underset{%
\mathcal{N}=2,D=4,n_{v}=3,n_{H}=4,~\text{self-mirror~}stu\text{~model}%
,~(B,F)=(32,32)}{\left[ \frac{SL(2,R)}{U(1)}\right] ^{\otimes 3}\times \frac{%
SO(4,4)}{SO(4)\times SO(4)}}.  \label{blues-2}
\end{equation}

\item The last step (\ref{jazz-3}) corresponds to the truncation of the
self-mirror $D=4$ $stu$ model to an $\mathcal{N}=1,D=4$ theory with $7$ WZ
multiplets, whose scalars span the \textit{seven-disk manifold} (\ref{0})
\cite{Duff-Ferrara-curious, Ferrara-Kallosh}:%
\begin{equation}
~\underset{\mathcal{N}=2,D=4,n_{v}=3,n_{H}=4,~\text{self-mirror~}stu\text{%
~model},~(B,F)=(32,32)}{\left[ \frac{SL(2,R)}{U(1)}\right] ^{\otimes
3}\times \frac{SO(4,4)}{SO(4)\times SO(4)}}~\longrightarrow \underset{%
\mathcal{N}=1,D=4,n_{c}=7,n_{v}=0,~(B,F)=(16,16)}{\left[ \frac{SL(2,R)}{U(1)}%
\right] ^{\otimes 7}}.~  \label{blues-3}
\end{equation}%
This step is non-trivial for what concerns the retaining of an $\mathcal{N}%
=1 $ local supersymmetry in the gravity theory with non-linear sigma model
given by (\ref{1}). Besides the necessary truncation of the $\mathcal{N}=1$
gravitino multiplet coming from the supersymmetric $\mathcal{N}=2\rightarrow
\mathcal{N}=1$ reduction of the $\mathcal{N}=2$ gravity multiplet, one has
to truncate all $\mathcal{N}=1$ vector multiplets coming from the
supersymmetry reduction of the three $\mathcal{N}=2$ vector multiplets;
furthermore, a truncation of half of the $\mathcal{N}=1$ chiral multiplets
stemming from the supersymmetry reduction of the four $\mathcal{N}=2$
hypermultiplets must be performed. This last step is particularly
challenging for the consistency with local $\mathcal{N}=1$ supersymmetry,
which is however granted by the results\footnote{%
In a different framework, more pertaining to the first path (\ref{2})-(\ref%
{2-tris}) to (\ref{1}), $\mathcal{N}=1$ local supersymmetry for the $D=4$
theory with scalar manifold (\ref{1}) was obtained in \cite%
{Duff-Ferrara-curious} by considering $M$-theory compactified on a suitable $%
7$-dimensional manifold with $G_{2}$-structure.} in \cite{ADF-SUSY-breaking}
(also \textit{cfr.} \cite{FKM-minimal-coupling}); see, in particular, the
discussion around Eq. (6.145) therein.
\end{enumerate}

\section{\label{Conclusion}Conclusion}

Summarizing, there exist (\textit{at least}) three different ways to obtain
the four $\mathcal{N}=16,8,4,2$ curious supergravities (\ref{128}), (\ref{64}%
), (\ref{32}) and (\ref{16}) with symmetric scalar manifolds of (maximal)
rank $8$ in $D=3$ :

\begin{enumerate}
\item Toroidal compactification of $M$-theory from $D=11$ to $D=3$,
respectively retaining geometric and non-geometric moduli of $T^{8}$, and
then geometric moduli of $T^{8}$, of $T^{4}\times T^{4}$, and of $%
T^{2}\times T^{2}\times T^{2}\times T^{2}$. This is given by the $M$\textit{%
-theoretical path} (\ref{2})-(\ref{2-tris}) discussed in Sec. \ref{M-Theory}.

\item Toroidal compactification of the four curious supergravities \cite%
{Duff-Ferrara-curious} (defined in $11,7,5,4$ dimensions) respectively on $%
T^{8}$, $T^{4}$, $T^{2}$, $T^{1}=S^{1}$ down to $D=3$; this is discussed at
point 3 of Sec. \ref{M-Theory}.

\item $S^{1}$-dimensional reduction $D=4\rightarrow D=3$ of the $\mathcal{N}%
=8,4,2,1$, $D=4$ curious supergravities with rank-$7$ scalar manifolds
(after dualization; \textit{cfr.} Table XVIII of \cite{Duff-Ferrara-curious}%
). This is given by the \textit{Ehlers path} (\ref{jazz-pre-2})-(\ref{jazz-4}%
) discussed in Sec. \ref{Ehlers}.\medskip
\end{enumerate}

By comparing the two paths (\ref{2})-(\ref{2-tris}) and (\ref{jazz-pre-2})-(%
\ref{jazz-4}), it is evident that they exhibit different and features.

The $M$\textit{-theoretical path} (\ref{2})-(\ref{2-tris}) is deeply rooted
in $M$-theory, and it makes \textit{\textquotedblleft octality"}, pertaining
to the symmetry of the fully factorised rank-$8$ Hodge-K\"{a}hler symmetric
coset (\textit{eight-disk manifold} (\ref{1})) in $D=3$, completely manifest
: the $SL(2,\mathbb{R})$'s of $T$-duality (from the $T^{2}$-factors of the $%
8 $-dimensional internal manifold), the $SL(2,\mathbb{R})$'s of $S$-duality
and $U$-duality, and the $D=4$ Ehlers group $SL(2,\mathbb{R})_{Ehlers}$ (of
gravitational origin) get \textit{unified}, and they stand on the same
footing.

On the other hand, the \textit{Ehlers path} (\ref{jazz-pre-2})-(\ref{jazz-4}%
), makes only \textit{\textquotedblleft septality"}, pertaining to the
full-fledged symmetry of the fully factorised rank-$7$ Hodge-K\"{a}hler
symmetric coset in $D=4$ (\textit{seven-disk manifold} (\ref{0})),
completely manifest : \textit{only} the $SL(2,\mathbb{R})$'s of $S$-, $T$-
and $U$- dualities get unified.

However, notwithstanding the first step (\ref{jazz-pre-2}) which seems to
single out the $D=4$ Ehlers group $SL(2,\mathbb{R})_{Ehlers}$, a \textit{%
complete equivalence} between the two paths is reached at their final steps.
It would be worth pursuing an $E_{11}$ interpretation \cite{West} of these
four maximal rank theories preserving minimal supersymmetry in $D=11,7,5,4$.

We also recall that in $D=4$ the four curious supergravities with $\mathcal{N%
}=8,4,2,1$ are associated with $7,3,1,0$ vertices of the Fano plane \cite%
{Duff-Ferrara-curious} . Similarly, in $D=3$ the $\mathcal{N}=16,8,4,2$
theories are associated with the $7,3,1,0$ quadrangles of the Fano plane and
the dual Fano plane\footnote{%
See the decompositions of the $\mathbf{56}$ under $E_{7}\supset \left[ SL(2)%
\right] ^{\otimes 7}$ and of the $\mathbf{248}$ under $E_{8}\supset \left[
SL(2)\right] ^{\otimes 8}$ in \cite{ICL-Rev}.}.

\section*{Acknowledgments}

We are grateful to Renata Kallosh, for useful discussions and related
collaboration. MJD is grateful to the Leverhulme Trust for an Emeritus
Fellowship and to Philip Candelas for hospitality at the Mathematical
Institute , Oxford. This work was supported by the STFC under rolling grant
ST/G000743/1. The work of SF is supported in part by CERN TH Dept. and
INFN-CSN4-GSS. AM wishes to thank the CERN Theory Division, for the kind
hospitality during the realization of this work. MJD and SF acknowledge the
hospitality of the GGI institute in Firenze, where this work was completed
during the workshop `Supergravity: What Next?'.

\end{document}